\documentclass{ptephy}

\usepackage{amsmath} 
\usepackage{amsthm} 
\usepackage{cite} 
\usepackage{graphicx} 
\usepackage{algorithmic} 
\usepackage{subfig} 
\usepackage{url} 
\usepackage{latexsym}
\usepackage{natbib}
\usepackage{bm}
\bibliography{sample}

\begin{document}
\title{Cosmological renormalization of model parameters in the second-order perturbation theory}
\author{\name{Kenji Tomita}{\ast}}
\address{\affil{}{Yukawa Institute for Theoretical Physics, Kyoto University, Kyoto 606-8502,
 Japan}
\rm{\email{ketomita@ybb.ne.jp}}}

\begin{abstract}
It is shown that the serious problem on the cosmological tension between the direct
 measurements of the Hubble constant at present and the constant derived from the 
Planck measurements of the CMB anisotropies can be solved by considering the 
renormalized model parameters. They are deduced by taking the spatial average of
 second-order perturbations in the flat $\Lambda$-CDM model, which includes
 random  adiabatic  fluctuations.

\end{abstract}

\maketitle

\section{Introduction}

High precision cosmology has started with the measurements of  fluctuations
in the cosmic microwave background radiation (CMB) by WMAP\citep{wmap} and
 Planck\citep{planck1, planck2} collaboration.
Their studies have been useful to determine the Hubble constant and the 
cosmological parameters.  However, it has been found that there is a tension 
between the Hubble constant ($H_0$) due to the Planck measurements of the CMB anisotropies and that due to the direct measurements.

Due to the Planck measurements\citep{planck1,planck2}, we have the Hubble constant
\begin{equation}
  \label{eq:k1}
H_0 = 67.3\ {\rm km \ s^{-1} Mpc^{-1}}.
\end{equation}
On the other hand, there are many direct measurements\citep{h4}, and the three
 examples of direct measurements give us\citep{h1,h2,h3}
\begin{equation}
  \label{eq:k2}
H_0 = 73.8, \ 74.3, \ {\rm{and}} \ 78.7 \ {\rm km \ s^{-1} Mpc^{-1}}.
\end{equation}
These data show that there is a large discrepancy of $9.7 \%  \sim 16.9 \%$.  
To solve this problem, various mechanisms have been proposed, such as 
models of decaying dark matter\citep{decay}, a local void model\citep{void}, and 
the dark radiation\citep{h4}. 
In this paper, it is shown that the tension can be solved by considering the
 renormalized 
model parameters which are deduced by taking  the spatial average of  
second-order perturbations in the flat $\Lambda$-CDM model, which includes 
random adiabatic fluctuations. 

In Sect. 2, we show the general-relativistic second-order perturbation theory
 in the flat 
$\Lambda$-CDM model, which was derived by the present author\citep{tom}.
In Sect. 3, we derive spatial averages of second-order density and metric
 perturbations,
and in Sect. 4, we define the renormalized Hubble constant due to the
average second-order metric perturbations, and show that it is consistent with 
the measured Hubble constants and their various observed values 
 correspond to the different upper limits of wave-numbers of
 perturbations which can be included in the renormalized perturbations.
Other renormalized model parameters are also derived due to the average of 
second-order density perturbations. 
In Sect. 5, the renormalization in the past is described and 
 in Sect. 6, the concluding remarks are given.  In Appendix A, we show the definition
 of various quantities included in the expressions for the second-order metric 
 perturbations. In Appendix B, we show the model parameters corresponding to 
 Hubble constants in Eq. (2).
 
\section{Background and the perturbation theory}

The background universe is expressed by a spatially flat model with the line-element
\begin{equation}
  \label{eq:a1}
ds^2 = g_{\mu \nu} dx^\mu dy^\nu = a^2 (\eta) [-d\eta^2 + \delta_{ij} dx^i dx^j ],
\end{equation}
where the Greek and Roman letters denote $0, 1, 2, 3$ and $1, 2, 3$, respectively. The conformal time $\eta (= x^0)$ is related to the cosmic time $t$ by $dt = a(\eta) d\eta$.

   In the comoving coordinates, the velocity vector and energy-momentum tensor of
pressureless matter are 
\begin{equation}
  \label{eq:a2}
u^0 = 1/a,  \quad u^i = 0
\end{equation}
and
\begin{equation}
  \label{eq:a3}
T^0_0 = -\rho, \quad T^0_i = 0, \quad T^i_j = 0,
\end{equation}
where $\rho$ is the matter density . 

From the Einstein equations, we obtain
\begin{equation}
  \label{eq:a4}
\rho a^2 = 3(a'/a)^2 - \Lambda a^2,
\end{equation}
and
\begin{equation}
  \label{eq:a5}
\rho a^3 = \rho_0,
\end{equation}
where a prime denotes $\partial/\partial \eta$,  $\Lambda$ is the
cosmological constant, and $\rho_0$ is an integration constant, and we use
the units $8\pi G = c = 1$ for the gravitational constant $G$ and the light velocity $c$. 
The Hubble parameter  $H$ is defined as 
\begin{equation}
  \label{eq:a6}
H  =  \dot{a}/a = a'/a^2.
\end{equation}
Eq.(\ref{eq:a4}) gives 
\begin{equation}
  \label{eq:a7}
H^2 = \frac{1}{3}  (\rho + \Lambda),
\end{equation}
which is also expressed as
\begin{equation}
  \label{eq:a8}
H^2 = H_0^2 \quad (\Omega_M a^{-3} + \Omega_{\Lambda} ),
\end{equation}
where $H_0$ is $H$ at the present epoch $t_0$ and $a_0 \equiv a(t_0) = 1$, and
\begin{equation}
  \label{eq:a9}
\Omega_M  = \frac{8\pi G \rho_0}{3H_0^2} = \frac{1}{3} \frac{\rho_0}{H_0^2}
  \quad {\rm{and}} \quad \Omega_{\Lambda} = \frac{\Lambda c^2}{3H_0^2} = \frac{1}{3} 
  \frac{\Lambda}{H_0^2}.
\end{equation}
In this paper we adopt the following background values : 
\begin{equation}
  \label{eq:a10}
H_0 = 67.3 \ {\rm km \ s^{-1} Mpc^{-1}} \  {\rm{and}} \
(\Omega_M,  \Omega_{\Lambda} )  = (0.22, 0.78).
\end{equation}
The significance of these values will be explained later.
   
Next let us show the first-order density perturbations. The perturbations of metric, matter
density and velocity are represented by $\delta_1 g_{\mu \nu} \ (\equiv h_{\mu\nu}),
\ \delta_1\rho,$ and $\delta_1 u^{\mu}$.   When we adopt the synchronous coordinates
(useful  in the pressureless case), the metric perturbations satisfy the condition
\begin{equation}
  \label{eq:a11}
h_{00} = 0 \quad {\rm{and}} \quad h_{0i} = 0.
\end{equation}
The first-order perturbations are classified into the growing case and the decaying case.
Both cases are found in the previous paper.\citep{tom} 
Here we show only those in the growing case, which are used in this paper :
\begin{equation}
  \label{eq:a12}
   \begin{split} 
h^j_i &= \delta^j_i F + P(\eta) F^{|j}_{|i}, \\
\mathop{\delta}_1 u^0 &= 0, \ \quad \mathop{\delta}_1 u^i = 0, \\
\mathop{\delta}_1 \rho/\rho &= \frac{1}{\rho a^2} \Bigl(\frac{a'}{a} P'
-1\Bigr) \Delta F, 
 \end{split} 
\end{equation}
where $F$ is an arbitrary function of spatial coordinates $x^1, x^2$
and $x^3$, \ $\Delta \equiv \nabla^2$, \ $h^j_i = g^{jl} h_{li}$ and
$P(\eta)$ satisfies  
\begin{equation}
  \label{eq:a13}
P'' + \frac{2a'}{a} P' - 1 =0.
\end{equation}
Its solution is expressed as
\begin{equation}
  \label{eq:a14}
 P = \int^\eta_0 d\eta' a^{-2} (\eta') \int^{\eta'}_0 d\eta'' a^2(\eta'').
\end{equation}
After a partial integration, we obtain 
\begin{equation}
  \label{eq:a15}
     \begin{split} 
(H_0)^2 \ P &= -\frac{2}{3\Omega_M} a^{-3/2} \sqrt{\Omega_M +\Omega_\Lambda 
a^3} \int^a_0
db \ {b}^{3/2}/\sqrt{\Omega_M +\Omega_\Lambda {b}^3} + \frac{2}{3\Omega_M} a,\\
H_0 \ \eta &= \int^a_0
db \ {b}^{-1/2}/\sqrt{\Omega_M +\Omega_\Lambda {b}^3}.
   \end{split} 
\end{equation}
The three-dimensional covariant derivatives $|i$ are defined in the
space with metric $dl^2 = \delta_{ij} dx^i dx^j$ and their suffices
are raised and lowered using $\delta_{ij}$, so that their derivatives
are equal to partial derivatives, i.e. $F^{|j}_{|i} = F_{,ij}$, where  
$F_{,i} \equiv  \partial F/\partial x^i$.

The second-order perturbations were derived in the previous
 paper\citep{tom}.  This is a simple extension of my paper\citep{tomold} 
which derived the second-order  perturbations in the case of
 zero $\Lambda$ in the
Lifshitz formalism with iterative second-order terms. The results in the 
latter paper were later derived independently by Russ et al.\citep{russ}
and by Matarrese et al.\citep{mat}
in the different formalisms, and the validity of this theory has been confirmed.
Here let us show their components 
$\mathop{\delta}_2 g_{\mu\nu} (\equiv \ell_{\mu\nu}), \mathop{\delta}_2
\rho,\ {\rm and} \ \mathop{\delta}_2 u^\mu $ in the case of
 nonzero $\Lambda$, where the total perturbations are 

\begin{equation}
  \label{eq:a16}
   \begin{split}
\delta g_{\mu\nu} &= h_{\mu\nu} + \ell_{\mu\nu}, \\
\delta u^\mu &= \mathop{\delta}_1 u^\mu +\mathop{\delta}_2 u^\mu, \\ 
\delta \rho/\rho &= \mathop{\delta}_1 \rho/\rho + \mathop{\delta}_2 \rho/\rho.
 \end{split}
\end{equation}
Here assuming the synchronous gauge condition, we have
\begin{equation}
  \label{eq:a17}
\ell_{00} = 0 \ \ {\rm and} \ \ \ell_{0i} = 0.
\end{equation}
The perturbations in the growing case are expressed as
\begin{equation}
  \label{eq:a18}
\ell^j_i = P(\eta) L^j_i + P^2 (\eta) M^j_i + Q(\eta) N^{|j}_{|i} + C^j_i,
\end{equation} 
where $N^{|j}_{|i} = \delta^{jl} N_{|li} = N_{,ij}$ and $Q(\eta)$ satisfies
\begin{equation}
  \label{eq:a19}
Q'' +\frac{2a'}{a} Q' = P - \frac{5}{2} (P')^2.
\end{equation} 
The expressions of other quantities $L^j_i ,  M^j_i , N^{|j}_{|i}, $ and $ C^j_i$ are
shown in Appendix A. 

The velocity and density perturbations are found to be
\begin{equation}
  \label{eq:a20}
\mathop{\delta}_2 u^0 =0, \quad  \quad \mathop{\delta}_2 u^i =0 
\end{equation} 
and
\begin{equation}
  \label{eq:a21}
     \begin{split}
\mathop{\delta}_2 \rho/\rho &= {1 \over 2\rho a^2}\Bigl\{{1 \over
2}(1 - \frac{a'}{a}P') (3F_{,l}F_{,l} + 8F\Delta F) +\frac{1}{2}P
[(\Delta F)^2 + F_{,kl}F_{,kl}] \\
 &+ \frac{1}{4}\Bigl[(P')^2 - \frac{2}{7} \frac{a'}{a}Q'\Bigr] [(\Delta F)^2 -
F_{,kl}F_{,kl}] - {1 \over 7} \frac{a'}{a}PP' [4F_{,kl}F_{,kl} +
3(\Delta F)^2]  \Big\}.
   \end{split}
\end{equation} 

The gauge used here is not only synchronous, but also comoving \ (cf. Eqs.
(\ref{eq:a12}) and (\ref{eq:a20})).  The above perturbations are therefore 
 physical density perturbations which are measured by comoving observers. 
 
 It is to be noticed that the present general-relativistic gravitational equations 
 are nonlinear and 
 applicable also in the super-horizon case, in contrast to linear gravitational 
 equations in the Newtonian treatment \citep{mak,bern},
 and so the cosmological result in the following
 sections cannot be derived in the Newtonian treatment, because of their difference. 
 
It is discussed in Sect. 6 what we should do, in order to obtain the 
consistency between the general-relativistic treatment and the 
Newtonian cosmology. 
\bigskip

\section{Average second-order perturbations}

We consider random perturbations given by
\begin{equation}
  \label{eq:b1}
F({\bf k})  = \int d{\bf k} \ \alpha ({\bf k}) \ e^{i{\bf kx}},
\end{equation}
where $\alpha ({\bf k}) $ is a random variable and the average of $F$ expressed as 
$\langle F \rangle $ vanishes. Here we assume the average process with a
power spectrum ${\cal P}_F ({\bf k})$, given by
\begin{equation}
  \label{eq:b2}
\langle \alpha ({\bf k}) \alpha ({\bf k'}) \rangle = (2\pi)^{-2}
{\cal P}_F ({\bf k}) \delta ({\bf k} + {\bf k'}).
\end{equation}
Here we have 
\begin{equation}
  \label{eq:b4}
\langle \delta_1 \rho/ \rho \rangle = 0
\end{equation}
for the first-order density perturbation. For the second-order  perturbations,
we have
\begin{equation}
  \label{eq:b5}
    \begin{split}
 F_{,i}  F_{,i} &= - \int \int d{\bf k}  d{\bf k'} 
\langle \alpha ({\bf k}) \alpha ({\bf k'}) \rangle {\bf k} {\bf k'} e^{i({\bf k}+
{\bf k'}) {\bf x}}, \\
 F \Delta F &= - \int \int d{\bf k}  d{\bf k'} 
\langle \alpha ({\bf k}) \alpha ({\bf k'}) \rangle k^2 e^{i({\bf k}+
{\bf k'}) {\bf x}},
  \end{split}
\end{equation}
so that we obtain
\begin{equation}
  \label{eq:b6}
\langle F_{,i}  F_{,i} \rangle  =  \langle F \Delta F \rangle = (2\pi)^{-2}  \int d{\bf k} \
k^2 {\cal P}_F ({\bf k}).
\end{equation}
Similarly, we have
\begin{equation}
  \label{eq:b7}
    \begin{split}
 F_{,ij}  F_{,ij} &=  \int \int d{\bf k}  d{\bf k'} 
\langle \alpha ({\bf k}) \alpha ({\bf k'}) \rangle ({\bf k} {\bf k'})^2 e^{i({\bf k}+
{\bf k'}) {\bf x}}, \\
 (\Delta F)^2 &=  \int \int d{\bf k}  d{\bf k'} 
\langle \alpha ({\bf k}) \alpha ({\bf k'}) \rangle k^2 (k')^2 e^{i({\bf k}+
{\bf k'}) {\bf x}},
  \end{split}
\end{equation}
so that we obtain
\begin{equation}
  \label{eq:b8}
\langle F_{,ij}  F_{,ij} \rangle  =  \langle (\Delta F)^2 \rangle = (2\pi)^{-2}  \int d{\bf k} \
k^4 {\cal P}_F ({\bf k}).
\end{equation}
\subsection{Second-order density perturbations}
It follows therefore from Eq. (\ref{eq:a21}) that
\begin{equation}
  \label{eq:b9}
\langle  \mathop{\delta}_2 \rho/\rho \rangle = 
\frac{1 - \frac{a'} {a}P'}{2\rho a^2} (2\pi)^{-2} \Big[\frac{11}{2} \int d{\bf k} k^2 
 {\cal P}_F ({\bf k}) + P \int  d {\bf k} k^4 {\cal P}_F ({\bf k})\Big].
\end{equation}
Here $F$ is related to the curvature fluctuation ${\cal R}$ by $F = 2\ {\cal R}$,
and so we have the relation 
\begin{equation}
  \label{eq:b10}
 {\cal P}_F  ({\bf k}) = 4\  {\cal P}_{\cal R}  ({\bf k}),
\end{equation}
where ${\cal P}_{\cal R}$ is  expressed using the power spectrum\citep{wein, tsuji}
  as
\begin{equation}
  \label{eq:b11}
 {\cal P}_{\cal R} = 2 \pi^2 \  {\cal P}_{{\cal R}0} \ k^{-3} (k/k_{eq})^{n-1} \ T_s^2
  (k/k_{eq})
\end{equation}
and $ {\cal P}_{{\cal R}0} = 2.2\times 10^{-9}$ according to the result of Planck
measurements.\citep{planck1,planck2}
The transfer function $T_s (x)$ is expressed as a function of $x = k/k_{eq}$, 
where 
\begin{equation}
  \label{eq:b11a}
k_{eq} \ (\equiv a_{eq} H_{eq}) \ = 219 \ (\Omega_M h) \ H_0 = 32.4 \ H_0. 
\end{equation}
Here $H_0$ \ ($\equiv 100h)$ \ is the present
 background  Hubble constant, \ $(a_{eq}, H_{eq})$ is $(a, H)$ at the epoch of
  equal energy density, and $(\Omega_M, h) = (0.22, 0.673)$ \ (given in
  Eq. (\ref{eq:a10})).

Moreover, we assume $n = 1$ here and in the following. 
Then we obtain for arbitrary $a$ 
\begin{equation}
  \label{eq:b12}
\langle  \mathop{\delta}_2 \rho/\tilde{\rho} \rangle = \frac{4\pi}{3} \ 32.4^4 \
 {\cal P}_{{\cal R}0} \ \frac{[1 - Y(a)]}{(\Omega_M/a + 
 \Omega_\Lambda a^2)} \Big[\frac{11}{2} \ 32.4^{-2} A + Z(a) B\Big], 
\end{equation}
where $\tilde{\rho} \equiv \rho + \Lambda$,  and $A$ and $B$ are expressed as 
\begin{equation}
  \label{eq:b13}
 A \equiv \int^{x_{max}}_{x_{min}} dx \ x \ T_s^2 (x),  \quad   B \equiv
  \int^{x_{max}}_{x_{min}}
 dx \ x^3 \ T_s^2 (x)
\end{equation}
using the transfer function $T_s (x)$ for the interval ($x_{max}, x_{min}$).  
Here we have
\begin{equation}
  \label{eq:b14}
Y(a) \ \equiv \frac{a'}{a}P', \quad Z(a) \ \equiv (H_0)^2 \ P,
\end{equation}
and using Eq.(\ref{eq:a15}), we obtain 
\begin{equation}
  \label{eq:b15}
Y(a) = a^{-5/2} (\Omega_M + \Omega_\Lambda a^3)^{1/2}  I(a), \quad 
Z(a) = \frac{2}{3\Omega_M} a [1 - Y(a)],
\end{equation}
where
\begin{equation}
  \label{eq:b17}
I(a) \equiv \int^a_0 \ db \ [b^3/(\Omega_M + \Omega_\Lambda b^3)]^{1/2}.
\end{equation}
\bigskip
\subsection{Second-order metric perturbations}
The second-order perturbation of the scale factor $\delta_2 a$ can be derived as
 follows using the metric second-order perturbations $l_{ij}$,  which are given 
in Eqs.(\ref{eq:a18}) and (\ref{eq:a19}), and in  Appendix A. 
The averaging of second-order metric perturbations leads to 
\begin{equation}
  \label{eq:sc1}
\delta_2 (a^2) = \frac{1}{3} \langle l^m_m  \rangle ,
 \end{equation} 
where we have
\begin{equation}
  \label{eq:sc2}
\langle l_{ij}  \rangle = P(\eta) \langle L_{ij}  \rangle + P^2 (\eta) \langle M_{ij}  \rangle
+ Q(\eta) \langle N_{,ij}  \rangle + \langle C_{ij}  \rangle.
 \end{equation} 
Since $L^i_j = L_{ij}, \ M^i_j = M_{ij}$ and 
\begin{equation}
  \label{eq:sc3}
    \begin{split}
L^i_i &= - \frac{1}{2} \ [2 F\Delta F + \frac{3}{2} F_{,l} F_{,l}], \\
M^i_i &= - \frac{1}{28} \ [10 (F_{,ll})^2 - 3 (\Delta F)^2 ],  \\
\Delta N &= - \frac{1}{28} \ [ (\Delta F)^2 -  F_{,kl} F_{,kl}], \\
\Box C^i_i &= 0,
\end{split}
 \end{equation} 
we obtain
\begin{equation}
  \label{eq:sc4}
    \begin{split}
\langle L^i_i  \rangle &= - \frac{7}{4} \ \langle F \Delta F  \rangle, \\
\langle M^i_i  \rangle &= - \frac{1}{4} \ \langle (\Delta F)^2  \rangle, \\
\langle \Delta N  \rangle &= \langle C^i_i  \rangle  = 0.
\end{split}
 \end{equation} 
Then we have using Eqs.(\ref{eq:b6}) and (\ref{eq:b8})
\begin{equation}
  \label{eq:sc5}
\langle l_{ii}  \rangle =  \langle l^i_i  \rangle  =  - 2\pi \ 32.4^4 \ {\cal P}_{R0} 
 \  Z(a) [7 \times 32.4^{-2} A + Z(a) B], 
 \end{equation} 
where $A, B$ and $Z(a)$ are given by Eqs.(\ref{eq:b13}) and (\ref{eq:b14}).

The line-element can be expressed as
\begin{equation}
  \label{eq:sc6}
ds^2 = -dt^2 + {a_{rem} (t)}^2 \ [(dx^1)^2 + (dx^2)^2 + (dx^3)^2],
 \end{equation} 
where the renormalized scale-factor is defined by
\begin{equation}
  \label{eq:sc7}
   {a_{rem} (t)}^2 \equiv  a^2 + \delta_2 (a^2) = a^2 + \frac{1}{3} \langle l^i_i  \rangle,
 \end{equation} 
and the relative difference of scale-factors is given by 
\begin{equation}
  \label{eq:sc7a}
   \xi \equiv  a_{rem} (t)/a(t) - 1 = \Big[1 + \frac{1}{3} \langle l^i_i 
    \rangle/a^2\Big]^{1/2} - 1.
 \end{equation} 

The renormalized redshift $z_{rem}$ corresponding to an arbitrary time $t$ is defined using the 
scale-factor $a_{rem}$ as
\begin{equation}
  \label{eq:sc8}
  1 +  z_{rem}  \equiv a_{rem} (t_0) / a_{rem} (t) = \frac{1+\xi (t_0)}{1+\xi (t)} (1+z),
 \end{equation} 
 where $t_0$ denotes the present epoch,  the background redshift $z$ is $1/a -1$, 
 and $\xi (t)$ is defined by Eq.(\ref{eq:sc7a}). 

The square of the background Hubble parameter $H$ is $(\dot{a}/a)^2$ and its
perturbation is given by
\begin{equation}
  \label{eq:sc9}
\delta_2 (H^2) = \delta_2 (\dot{a}^2)/a^2 - (\dot{a})^2 \delta_2 (a^2)/a^4,
 \end{equation} 
so that
\begin{equation}
  \label{eq:sc10}
\frac{\delta_{2} (H^2)}{H^2} = \frac{2}{3} \Big[\frac{\langle l^m_m  \rangle'}{2aa'} - 
\frac{\langle l^m_m  \rangle}{a^2}\Big].
 \end{equation} 
From Eqs.(\ref{eq:sc5}) and (\ref{eq:sc10}), we obtain
\begin{equation}
  \label{eq:sc11}
\frac{\delta_{2} (H^2)}{H^2} = \frac{4\pi}{3} 32.4^4  {\cal P}_{R0}  \
\frac{1}{a^2} \Big[7 (32.4)^{-2} \Big(Z - \frac{Y(a) a}{2(\Omega_M + \Omega_\Lambda
 a^3)}\Big)
A + Z \Big(Z - \frac{Y(a) a}{\Omega_M + \Omega_\Lambda a^3}\Big) B\Big],
 \end{equation} 
where $Y(a)$ and $Z(a)$ are given by Eq.(\ref{eq:b15}).

\bigskip
\subsection{Average perturbations of model parameters}
In this paper we assume the simplest transfer function (BBKS) for cold matter,
adiabatic fluctuations,  given by\citep{bbks,ll,dodel}
\begin{equation}
  \label{eq:b17a}
T_s (x) = \frac{\ln (1+0.171 x)}{0.171x} [1+0.284 x +(1.18 x)^2 + (0.399 x)^3 +
(0.490 x)^4]^{-1/4}.
\end{equation}
This function has the peak around  $x \simeq 1$, so that the upper and lower limits
 ($x_{max}$ and $x_{min}$) in the integrals $A$ and $B$ in Eq. (\ref{eq:b12}) should 
 have the values 
 such as $x_{max} \sim 6 (> 1)$ and $x_{min} \sim 0.01 (\ll 1).$
Here  $A$ and $B$ depend sensitively on $x_{max}$, but not on $x_{min}$.

In order to find the best value of $x_{max}$, we derive a length $L_{max}$ 
corresponding to $k_{max}$. Using Eq.(\ref{eq:b11a}), we have     
\begin{equation}
  \label{eq:b17a1}
L_{max} \equiv 2\pi/k_{max} = 102/h  \ {\rm{Mpc}}  
\end{equation}
for $x_{max} \ (\equiv k_{max}/k_{eq}) = 5.7$.  This $L_{max}$ corresponds to
the cosmological distance, over which the smooth  observations on
cosmological scales may be possible.  

So we adopt $x_{max} = 5.7$ and $x_{min} = 0.01$. Then we obtain
\begin{equation}
  \label{eq:b17b}
A = 2.22, \ \quad \ B = 20.95,
\end{equation}
and 
\begin{equation}
  \label{eq:b17b1}
\langle  \mathop{\delta}_2 \rho/\tilde{\rho} \rangle = 0.121 \quad  {\rm{and}} \quad
\frac{\delta_2 (H^2)}{H^2} = 0.210 
\end{equation}
at the present epoch ($ a = 1$).

  In similar cases with $B \approx 10 A$,
 the terms with $A$ in Eqs. (\ref{eq:b12}) and (\ref{eq:sc11}) are negligible, and then we have 
\begin{equation}
  \label{eq:b17c}
\langle  \mathop{\delta}_2 \rho/\tilde{\rho} \rangle \simeq \frac{8\pi}{9} 32.4^4 \
 {\cal P}_{{\cal R}0}  \ \frac{[1-Y(a)]^2 a^2/\Omega_M}{\Omega_M + \Omega_\Lambda 
 a^3} \ B ,
\end{equation}
and
\begin{equation}
  \label{eq:b17c1}
\frac{\delta_2 (H^2)}{H^2} \simeq  \frac{4\pi}{3} 32.4^4 \ {\cal P}_{{\cal R}0} 
\frac{1}{a^2}  Z(a) \Big[Z(a) - \frac{Y(a) a}{\Omega_M + \Omega_\Lambda a^3}\Big] B.
\end{equation}

By the way, we derive $(\delta_1 \rho/\tilde{\rho})^2$ to estimate the dispersion of 
$\langle  \mathop{\delta}_2 \rho/\tilde{\rho} \rangle$:
\begin{equation}
  \label{eq:b18}
\langle  (\mathop{\delta}_1 \rho/\rho)^2 \rangle = 
\Big[\frac{1 - Y(a)}{\rho a^2}\Big]^2 (2\pi)^{-2}   \int  d {\bf k} k^4 {\cal P}_F ({\bf k}),
\end{equation}
so that we obtain
\begin{equation}
  \label{eq:b19}
\langle  (\mathop{\delta}_1 \rho/\tilde{\rho})^2 \rangle = \frac{8\pi}{9} 
\times 32.4^4 \
 {\cal P}_{R0} \ \frac{[1-Y(a)]^2 a^2}{(\Omega_M + \Omega_\Lambda a^3)^2} 
 B \simeq \langle  \mathop{\delta}_2 \rho/\tilde{\rho} \rangle \frac{\Omega_M}{\Omega_M +\Omega_\Lambda a^3}.
\end{equation}
%
\bigskip

\section{Renormalization of model parameters}

Now let us consider the renormalization of the background density and the Hubble
constant. 
Since $\langle  \mathop{\delta}_2 \rho/\rho \rangle$ is spatially constant, 
we may 
assume that it is a part of the background density. Here we regard 
\begin{equation}
  \label{eq:c0a}
\rho_{rem} \equiv \rho + \langle  \mathop{\delta}_2 \rho \rangle
\end{equation}
as the renormalized background density.

Similarly $\langle \delta_2 H^2 \rangle$ is spatially constant, and so we can regard 
\begin{equation}
  \label{eq:c1}
H_{rem} \equiv [H^2 +  \langle \delta_2 (H^2) \rangle]^{1/2}
\end{equation}
as the renormalized background Hubble parameter. 

For  $\langle \mathop{\delta}_2  H^2  \rangle/H^2$ in Eq. (\ref{eq:b17b1}) and 
$ H = H_0$ in Eq. (\ref{eq:a10}), we obtain at present epoch
\begin{equation}
  \label{eq:c2a}
H_{rem} = 74.0 \ \rm{km \ s^{-1}\ Mpc^{-1}} ,
\end{equation}
which is equal approximately 
 to the measured Hubble constants.\citep{h2,h3}  It is found, therefore, that the
renormalized Hubble constant $H_{rem}$ may be consistent with the directly measured 
Hubble constants.

The model parameters $\Omega_M$ and $\Omega_\Lambda$ describe the
evolution of the background universe. But since our real universe is  described using
the renormalized quantities $H_{rem}$ and $\rho + \langle 
   \mathop{\delta}_2 \rho \rangle$ in the place of the
   background Hubble constant $H$ and $\rho$, we may obtain the
   following new set of model parameters :
\begin{equation}
  \label{eq:c9}
({\Omega}_M)_{rem} = \Omega_M \frac{1 +  \langle  \mathop{\delta}_2
 \rho/\rho  \rangle}{1 + \langle  \mathop{\delta}_2 \rho/\tilde{\rho}  \rangle}
\end{equation}
and
\begin{equation}
  \label{eq:c10} 
({\Omega}_\Lambda)_{rem}  = \Omega_\Lambda  \frac{1}{1 + \langle  \mathop{\delta}_2
 \rho/\tilde{\rho}  \rangle}.
\end{equation}
Using the background model parameters Eq.(\ref{eq:a10}) and $\langle 
 \mathop{\delta}_2 \rho/\tilde{\rho} \rangle$ in Eq. (\ref{eq:b17b1}), 
we obtain at the present epoch
\begin{equation}
  \label{eq:c11} 
({\Omega}_M)_{rem} = 0.305 \ \quad  {\rm{and}} \quad \
 ({\Omega}_\Lambda)_{rem} = 0.695.
\end{equation}

The recent observations of the redshift-magnitude relation\citep{betoule} 
include many supernova
 with redshifts $z = 0.1 - 1.0$, so that the present Hubble constant ($H_{rem}$)
is used, and  $(({\Omega}_M)_{rem},  ({\Omega}_\Lambda)_{rem})$ \ (which are
 consistent with Eq.(\ref{eq:c11})) are obtained.

As for  the observations of baryon acoustic oscillations of CMB
 (Planck\citep{planck1,planck2}) and large-scale galactic correlations\citep{eisen,perc},
we use the angular distance in the late time model, so that the derived model
 parameters  are
not $({\Omega}_M,  {\Omega}_\Lambda)$, but $(({\Omega}_M)_{rem}, 
 ({\Omega}_\Lambda)_{rem})$,  which are given by Eq.(\ref{eq:c11}). On the other hand, 
 the scale of acoustic oscillations is determined at the recombination epoch with
 $a  (\approx 10^{-3}) $ and so the Hubble constant is given by $H_0$ \ ($\simeq
 H_{rem}$ at the recombination epoch), but not $H_{rem}$ at the present epoch.
 So the above renormalized model parameters are consistent with the cosmological 
 observations. \citep{planck1,planck2, eisen,perc}

The relative difference of scale-factors $\xi$ (in Eq.({\ref{eq:sc7a})) is $-0.097$ at
 present epoch. Moreover, the present values of $\langle  \mathop{\delta}_2
  \rho/\tilde{\rho}  \rangle$ and 
 $((\Omega_M)_{rem},   ({\Omega}_\Lambda)_{rem})$
in the case when $ H_{rem} $ is $73.8, 74.3$ or $78.7$ are shown in Appendix B.

\bigskip
\section{Renormalization of model parameters in the past}

In the previous section, we treated the quantities at the present epoch ($a = 1$).
Here we consider the quantities at the epochs of $a < 1$. First we calculate 
 $\langle  \mathop{\delta}_2 \rho/\tilde{\rho} \rangle$ for $a < 1$ 
 using Eq.(\ref{eq:b12}) for
  $x_{max} = 5.7$ and $x_{min} = 0.01$. Its dependence on $a$ is shown in Fig.1.
  It is found that $\langle  \mathop{\delta}_2 \rho/\tilde{\rho} \rangle$ has a peak at 
 around $a \sim 0.65$, but $\langle  \mathop{\delta}_2 \rho/{\rho} \rangle$ increases
  monotonically,  and that 
$\langle  \mathop{\delta}_2 \rho/\tilde{\rho} \rangle$ reduces to $0$ in the limit of $a 
\rightarrow 0$. 

  Using Eqs. (\ref{eq:sc11}),  (\ref{eq:c1}), (\ref{eq:c9}), and (\ref{eq:c10}), 
  moreover, we obtain 
$H_{rem}$ and $(({\Omega}_M)_{rem},   ({\Omega}_\Lambda)_{rem})$ in the past. 
They are shown in Fig. 2 and Fig. 3.
It is found that $H_{rem}$ reduces to $H_0$ \ (in Eq. (\ref{eq:a10})), and  
$(({\Omega}_M)_{rem},   ({\Omega}_\Lambda)_{rem})$ reduces to $({\Omega}_M,  
 {\Omega}_\Lambda)$ in the limit of $a \rightarrow 0$. 
 
 In Fig. 4, the relative difference of scale-factors $\xi$ (in Eq.(\ref{eq:sc7a})) is
 shown and $\xi$ is  $-0.097 \sim -0.195$ for $a = 1 \sim 0$, respectively. 
 It is found from Eq.(\ref{eq:sc8}) that $z_{rem}$ is larger than $z$.

\begin{figure}
\caption{\label{fig:1} The second-order perturbations $\langle  \mathop{\delta}_2
 \rho/\tilde{\rho}  \rangle$ is expressed as a function of $a$. \ The scale 
 factor $a$ has $1$ at the present epoch.}
\centerline{\includegraphics[width=10cm]{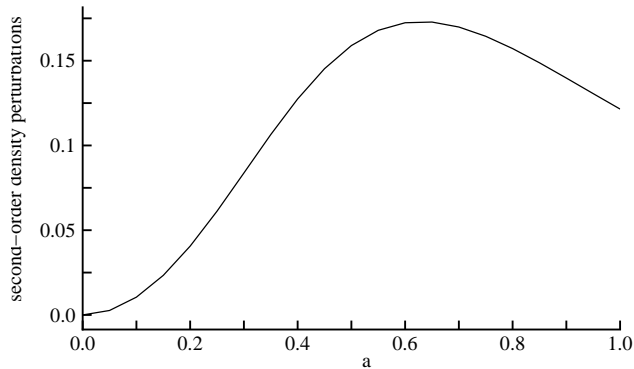}}
\end{figure}
\begin{figure}
\caption{\label{fig:2} The renormalized Hubble constant $H_{rem}$ is expressed 
as a function of $a$. \ The scale 
 factor $a$ has $1$ at the present epoch.}
\centerline{\includegraphics[width=10cm]{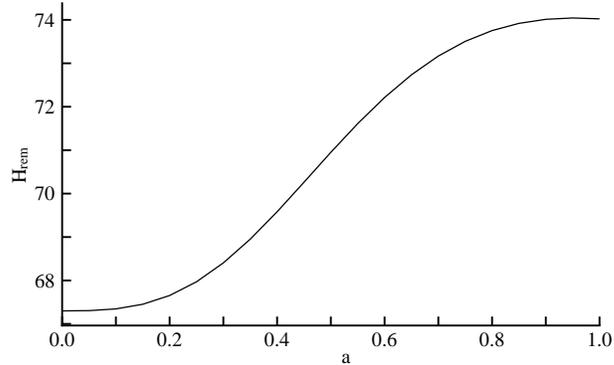}}
\end{figure}
\begin{figure}
\caption{\label{fig:3} The renormalized density parameter  $({\Omega}_M)_{rem}$ is
 expressed as a function of $a$. \ $({\Omega}_\Lambda)_{rem} = 1 -
 ({\Omega}_M)_{rem}$. \ The scale 
 factor $a$ has $1$ at the present epoch.}
\centerline{\includegraphics[width=10cm]{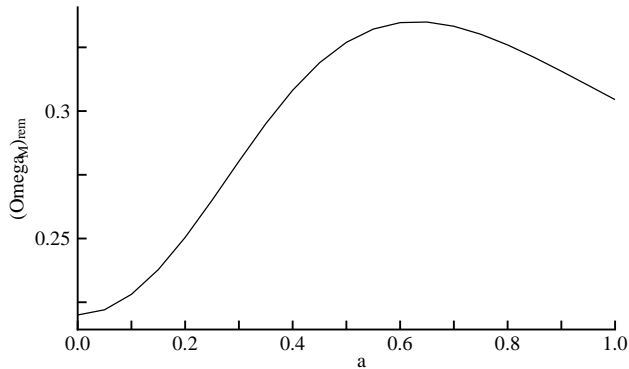}}
\end{figure} 
\begin{figure}
\caption{\label{fig:4} The relative difference of scale-factors   $\xi (t) \ (\equiv    
a_{rem}(t)/a(t) - 1)$ \  is expressed as a function of $a$.  \ The scale 
 factor $a$ has $1$ at the present epoch.}
\centerline{\includegraphics[width=10cm]{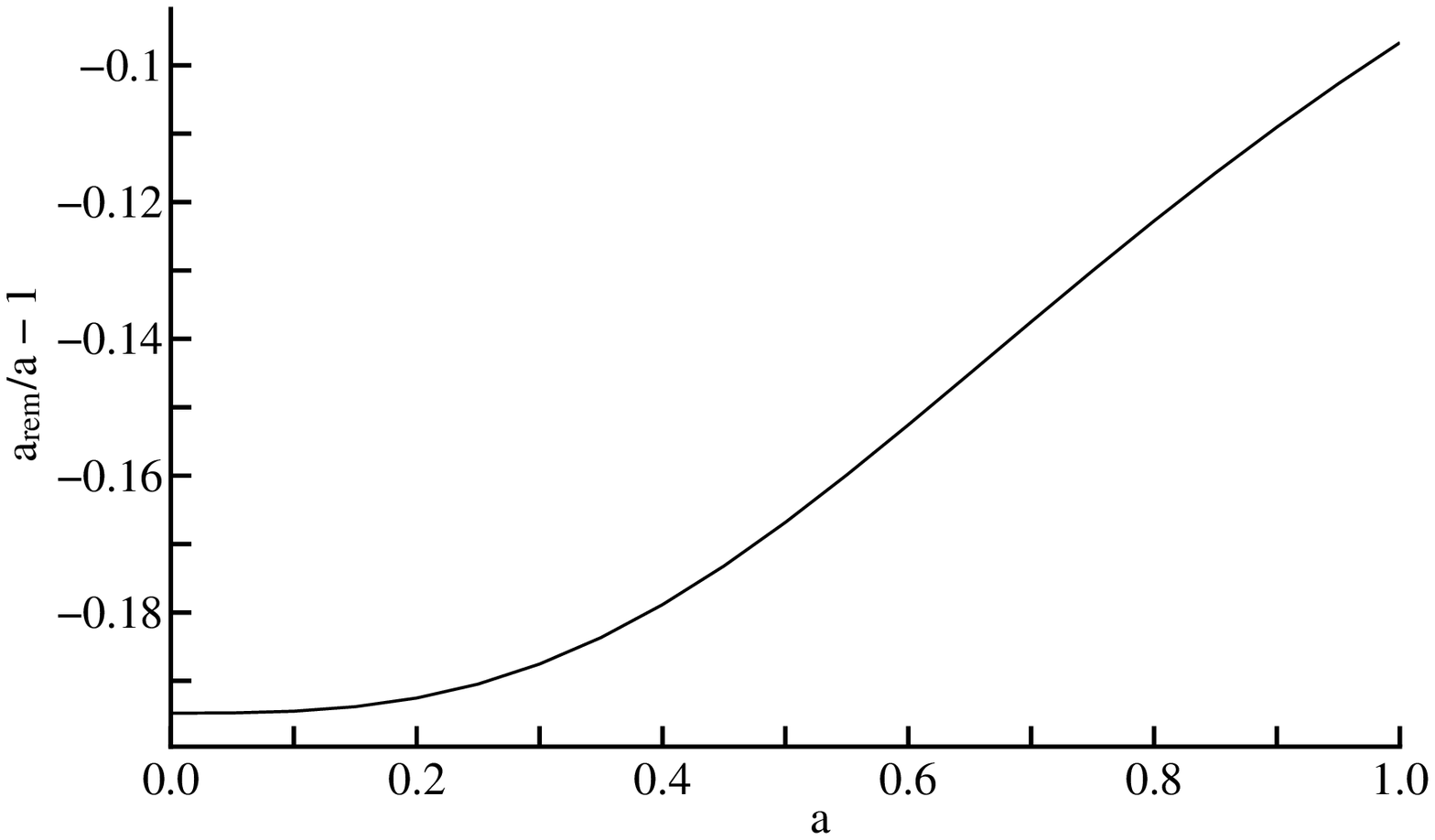}}
\end{figure}  
 
\bigskip

\bigskip
\section{Concluding remarks}
It was found in this paper that the random adiabatic fluctuations bring a kind
of energy density which has an influence upon the dynamics of the universe. 
For its derivation, the nonlinearity of general-relativistic perturbation theory was
 important.

As $k$ increases in the region of $x (\equiv k/k_{eq}) > 1$, the amplitude of
 perturbations decreases rapidly,  and the frequency of perturbed objects is so small
 that they cannot be renormalized as part of the background matter density.
 The upper limit of $x$ for renormalized perturbations is  $x_{max} (\approx 6)$.
 Because of their small frequency corresponding to large $k$, the value of $x_{max}$
  has large fluctuations, 
 and this may be the origin of the directional fluctuations of $x_{max}$ and the
 measured Hubble constant.

The background model parameters in Eq.(\ref{eq:a10}) are rather different from the
renormalized parameters in Eqs.(\ref{eq:c2a}) and (\ref{eq:c11}). We should notice
that the observed model parameters are the latter ones.  The Hubble constant $H_0$
 in the Planck measurements (Eq.(1)) is the renormalized Hubble constant 
 measured at the early stage  ($a \ll 1$), which is approximately equal to the
 background  Hubble constant ($H_0$).

 In this paper we adopted tentatively the background model parameters 
 (in Eq. (\ref{eq:a10})) and  the value of  $x_{max}$. Their values should be
  selected
 so that they may  reflect best the real observations of model parameters.
 
Large-scale perturbations such as $x \equiv k/k_{eq} = 0.01 \sim 6$, 
which were treated in this paper (cf. Eq.(\ref{eq:b17a1})) cross the 
horizon in the course of their evolution, so that taking the 
general-relativistic effect into consideration is indispensable for their 
dynamical analyses, which are not only linear but also on second-order.
 
In the Newtonian theory, the terms representing the gravitational strength 
$\xi \ (\equiv \ GM/(c^2 R) \ )$ are taken only linearly into account,
 assuming that it is exremely small,
where $M$ and $R$ are characteristic mass and length of dynamical 
objects.  In the cosmological circumstances, however, we have  
\begin{equation}
  \label{eq:ap1}
\xi \sim G \rho /H^2  \cdot (\delta \rho/ \rho) \ \sim \ \delta \rho/ \rho,
\end{equation} 
where $H$ is the Hubble parameter, $R \sim c/H$, and $M \ \sim \ 
\rho R^3$. When 
 $| \delta \rho/\rho|$ is not so small, we cannot neglect the
  nonzero mean of $\xi^2$, which should be treated
 in the post-Newtonian approximation.  In the general-relativistic treatment
 of second-order perturbations,
  this nonzero mean is automatically taken into account.
 In the linear level, we have $\langle \xi  \rangle = 0$,
 but in the second-order,  $ \langle \xi^2  \rangle \sim
\langle (\delta \rho/ \rho)^2 \rangle \ne 0$.
 Moreover, the necessity of considering nonlinear
 $\xi$ comes also from the energy-momentum conservation law.
In order that the Newtonian theory is compatible with the 
energy-momentum conservation in the same way as in the 
general-relativistic
 cosmology, we must add a nonzero term $(\delta \rho/\rho)^2$ 
to the second-order density perturbation,  so as to recover the 
post-Newtonian terms with $\xi^2 $. By this addition, 
the Newtonian theory may be consistent with the general-relativistic
 cosmology. The correspondence of general-relativistic approach and 
 Newtonian approach has been studied by Matarrese et al.\citep{mat2} 
 The discussions about it in detail are beyond the scope of the present 
 paper.

\bigskip
\section*{Acknowledgements}
The author thanks Masumi Kasai for helpful discussions. 
\bigskip

\appendix
\section{Quantities in the second-order metric perturbation}
 
The quantity $Q$  in the second-order metric perturbation is expressed as
\begin{equation}
  \label{eq:m1}
Q = \int^\eta_0 d\eta' a^{-2}(\eta') \int^{\eta'}_0 d\eta''
a^2(\eta'') \Bigl[P(\eta'') -{5 \over 2}(P'(\eta''))^2\Bigr]. 
\end{equation} 
In the case $\Lambda = 0$, we have \ $P - (5/2)(P')^2 = 0$ because of $a 
\propto \eta^2$ and $P = \eta^2/10$, so that $Q$ vanishes.  
The functions $L^j_i$ and $M^j_i$ are defined by
\begin{equation}
  \label{eq:m2}
    \begin{split}
    L^j_i &= {1 \over 2}\Bigl[-3 F_{,i} F_{,j} -2 F \cdot F_{,ij} + {1 \over 2}
\delta_{ij} F_{,l} F_{,l}\Bigr], \\
M^j_i &= {1 \over 28}\Big\{19F_{,il} F_{,jl} - 12 F_{,ij} \Delta F -
3\delta_{ij} \Bigl[F_{,kl} F_{,kl} -(\Delta F)^2 \Bigr]\Big\}
 \end{split}
\end{equation} 
and $N$ is defined by
\begin{equation}
  \label{eq:m3}
\Delta N = {1 \over 28} \Bigl[(\Delta F)^2 - F_{,kl}F_{,kl}\Bigr].
\end{equation} 

The last term $C^j_i$ satisfies the wave equation
\begin{equation}
  \label{eq:m4}
\Box C^j_i = {3 \over 14}(P/a)^2 G^j_i + {1 \over 7}\Bigl[P - {5 \over 2}
(P')^2 \Bigr] \tilde{G}^j_i,
\end{equation} 
where the operator $\Box$ is defined by
\begin{equation}
  \label{eq:m5}
\Box \phi \equiv g^{\mu\nu} \phi_{;\mu\nu} = -a^{-2}
\Bigl(\partial^2/\partial\eta^2 + {2a' \over a}\partial/\partial \eta -
\Delta \Bigr) \phi
\end{equation} 
for an arbitrary function $\phi$ by use of the four-dimensional covariant
derivative $;$, and $G^j_i$ and $\tilde{G}^j_i$ are expressed as 
\begin{equation}
  \label{eq:m6}
  \begin{split}
G^j_i &\equiv \Delta(F_{,ij} \Delta F - F_{,il} F_{,jl}) + (F_{,ij} F_{,kl}
- F_{,ik} F_{,jl})_{,kl} - {1 \over 2}\delta_{ij}\Delta [(\Delta F)^2 -
F_{,kl}F_{,kl}], \\
\tilde{G}^j_i &\equiv F_{,ij} \Delta F - F_{,il} F_{,jl} - {1 \over
4}\delta_{ij}[(\Delta F)^2 - F_{,kl}F_{,kl}] - 7 N_{,ij}. 
 \end{split}
\end{equation}
These functions satisfy the traceless and transverse relations
\begin{equation}
  \label{eq:m7}
    \begin{split}
G^l_l = 0, \quad G^l_{i,l} = 0, \\
\tilde{G}^l_l = 0, \quad \tilde{G}^l_{i,l} = 0,
  \end{split}
\end{equation}
so that $C^j_i$ also satisfies
\begin{equation}
  \label{eq:m8}
C^l_l = 0, \quad C^l_{i,l} = 0.
\end{equation} 
This means that $C^j_i$ represents the second-order gravitational
radiation emitted by first-order density perturbations.
The solution of the above inhomogeneous wave equation
(Eq.(\ref{eq:m4})) can be represented in an explicit form using 
the retarded Green function for the operator $\Box$.
\bigskip

\section{Case when $ H_{rem} $ is $73.8, 74.3,$ or $78.7$}
If we assume that the three values of  Hubble constant due to direct measurements
 $(= 73.8, 74.3, 78.7)$ are $H_{rem}$, we obtain
\begin{equation}
  \label{eq:c3}
\langle  \mathop{\delta}_2 \rho/\tilde{\rho}  \rangle = 0.117, \ 0.126, \ 0.213,
\end{equation}
respectively, and
\begin{equation}
  \label{eq:c4}
\langle  (\mathop{\delta}_1 \rho/\tilde{\rho} )^2 \rangle^{1/2} = 0.160, \ 0.166,
\ 0.216.
\end{equation}
Moreover,  corresponding to $H_{rem} = 73.8, 74.3, 78.7$ and $ H = H_0$ in Eq.
 (\ref{eq:a10}), it is found 
for a fixed $x_{min} = 0.01$ using Eq. (\ref{eq:b12}) that 
\begin{equation}
  \label{eq:c6}
 x_{max} = 5.6, \ 5.8, \ 7.4,
\end{equation}
\begin{equation}
  \label{eq:c7}
A = 2.20, \ 2.25, \ 2.59, \quad  B = 20.14, \ 21.78, \ 36.65,
\end{equation}
respectively.
The values of $A$ and $B$ depend rather sharply on $x_{max}$, but not on
 $x_{min}$. It is found therefore that the directly observed Hubble constants 
 appear, corresponding to various values of $x_{max} \ (\approx 6)$. 
 These values of $x_{max}$ may depend on the direction in which we measure
 the Hubble constant.

For  $H_{rem} = 73.8, 74.3, 78.7$, moreover, we obtain 
%
\begin{equation}
  \label{eq:c12} 
({\Omega}_M)_{rem} = (0.302, \ 0.307, \ 0.357) \quad  {\rm{and}} \quad 
 ({\Omega}_\Lambda)_{rem} = (0.698, \ 0.693, \ 0.643),
\end{equation}
respectively.

\bigskip


\end{document}